\documentclass{./aa}   %
\usepackage{graphicx}

\begin{document} %
\def\first{1$\rm^{st}$~}  
\def\second{2$\rm^{nd}$~}  
\def\third{3$\rm^{rd}$~}
\def\fourth{4$\rm^{th}$~}  
\def\irascolor{$\Delta I_{60}$/$\Delta I_{100}$}
\newcommand{\lele}[3]{{#1}\,$\le$\,{#2}\,$\le$\,{#3}}
\title{Catalogue of far-infrared loops in the Galaxy}

\author{ V. K\"onyves\inst{1}, Cs. Kiss\inst{2}, A. Mo\'or\inst{2},  Z.
     Kiss\inst{3} \and L.V. T\'oth\inst{1}    }

\offprints{V. K\"onyves, v.konyves@astro.elte.hu}

\institute{Astronomy Department, E\"otv\"os Lor\'and University,   P.O. Box 32,
H-1518 Budapest, Hungary\\  \email{v.konyves@astro.elte.hu,
l.v.toth@astro.elte.hu}  
\and Konkoly Observatory of the Hungarian Academy of
Sciences,  P.O. Box 67, H-1525~Budapest, Hungary\\ 
\email{pkisscs@konkoly.hu,moor@konkoly.hu}  
\and Baja Astronomical Observatory of
B\'acs-Kiskun County,  P.O. Box 766, H-6500~Baja, Hungary     
\newline
\email{kissz@alcyone.bajaobs.hu} } %

\date{Received; accepted} %

\abstract{}{An all-sky survey of loop- and arc-like intensity enhancements has
been performed in order to investigate the large-scale structure of the diffuse
far-infrared emission.}  
{We used maps made of 60 and 100\,$\mu$m 
processed IRAS data (Sky Survey Atlas and dust infrared emission maps) 
to identify large-scale structures: loops, arcs or cavities, 
in the far-infrared emission in the Galaxy.
Distances were attributed to a subsample of  
loops using associated objects.}  
{We identified 462 far-infrared loops, analyzed their individual FIR
properties and  their distribution. This data forms the Catalogue of 
Far-Infrared Loops in the Galaxy. We obtained observational estimates of 
$f_{\mathrm{in}}$\,$\approx$\,30\%
and $f_{\mathrm{out}}$\,$\approx$\,5\% for the hot gas volume filling 
factor of the inward and outward Galactic neighbourhood of the Solar System. 
We obtained a
slope of the power law size luminosity function $\beta$\,=\,1.37 for low
Galactic latitudes in the outer Milky Way. }  
{Deviations in the celestial
distribution of far-infrared loops clearly indicate, that violent events 
frequently overwrite the structure of the interstellar matter in the inner
Galaxy. Our objects trace out the spiral arm structure  of the Galaxy in the
neighbourhood of the Sun and  their distribution clearly suggests that  there is
an efficient process that can generate loop-like features at high  Galactic
latitudes. Power law indices of size luminosity distributions suggest, that the
structure of the ISM is ruled by supernovae and stellar winds at low Galactic
latitudes while it is governed by supersonic turbulence above the Galactic
plane.} 

\maketitle 


\section{Introduction}

The large scale structure of the cold interstellar matter  can be significantly
affected by violent events.  This structure -- which is diverse with the complex
distribution  of shells, cavities, filaments, arcs and loops -- is often referred
to  as the "Cosmic Bubble Bath" (Brand \& Zealey \cite{Brand75}). The evolution
of bubbles and superbubbles produced by supernova explosions and stellar winds of
associations are the primary processes that determine the structure and
energetics of all components of the diffuse interstellar medium. Studying
supernova (SN) explosions in a uniform medium, Cox and Smith \cite{Cox74} pointed
out that if the Galactic SN  rate is "sufficient" it can produce the "swiss
cheese morphology" of  the cold diffuse ISM with hot coronal gas inside the
bubbles. This was built into the model by McKee and Ostriker \cite{McKee77} where
SN explosions in the cloudy ISM produce a three-phase medium.  In this scenario
the next generation of stars are born in the compressed medium of the bubble
walls (propagating star formation, see Blaauw \cite{Blaauw91}, for a review).  

The idea that supersonic turbulence may be important in the control of star
formation became recognized in recent years 
(see MacLow \& Klessen \cite{MacLow04}, for a review). 
Turbulence -- if dominant --  can also rule the structure of
the ISM creating clouds and cavities with regular properties of a fractal
geometry, apparently similar to the structure formed via violent events. 

Structures produced by various processes are often characterized by the power 
law parameter $\beta$ of the mechanical luminosity distribution of bubbles or holes. 
With a specific model this can be converted into a parameter describing 
the size distribution of these structures, then $\beta$ can be determined even 
for a medium with no stellar energy injection (e.g. a structure created by supersonic 
turbulence).
Oey \& Clarke \cite{Oey97} obtained 1.4\,$\leq$\,$\beta$\,$\leq$\,2.1 
for the nearby galaxies M31, M33, Holmberg\,II, SMC. 
Ehlerov\'a and Palou\v{s} (\cite{Ehlerova05}) 
deduced $\beta$=1.6$\pm$0.3 from the size distribution of HI 
shells in the outer Milky Way. Elmegreen (\cite{Elmegreen99}) 
derived $\beta$\,=\,2.15 for a medium with a structure of $D$\,=\,2.3 fractal 
dimension, also from the size distribution of holes. This structure should be 
representative of that created by supersonic turbulence.   

The hot gas volume filling factor $f$ of a galaxy is also an important parameter
in characterizing the lifecycle  of the interstellar medium. Although there are
several estimates  of $f$ based on theoretical considerations and semi-empirical
models, the $f$ parameter of our Galaxy is hardly constrained observationally
(Ferri\'ere \cite{Ferriere98} and Gazol-Pati\~no \& Passot
\cite{Gazol-Patino99}). Recently Ehlerov\'a and Palou\v{s} (\cite{Ehlerova05})
obtained $f$=5\% for the shell volume filling factor for the outer Galaxy
through an automated identification of HI shells.  

To derive these global parameters a large sample of 
object are needed, and it is not easy to construct such a database
for our own Galaxy. 
Thilker et al. \cite{Thilker98} and Mashchenko et al. \cite{Mashchenko99} made
efforts to detect HI shells automatically, based on a model of the shell. 
Recently Ehlerov\'a et al. (\cite{Ehlerova04}) developed a model-independent 
algorithm which automatically searches HI shells in data cubes and identified
$\sim$1000 structures in the Leiden-Dwingeloo Survey data. An artificial neural
networks algorithm by Daigle et al. \cite{Daigle03} was successfully applied to
the Canadian Galactic Plane Survey data. However, these works intended to detect
HI holes based on the velocity information of the 21\,cm line data rather than on
morphology, therefore they cannot find bubbles at late stages of their evolution
when the expansion has slowed down.

Shell- or arc-like intensity enhancements are reported in many tracers of  the
ISM (see Kiss et al. \cite{Kiss04}, and references therein, for an
introduction). The structure of the diffuse ISM is well-represented in the far-infrared,
mainly observed as the Galactic cirrus emission (Low et al. \cite{Low84}).
Although some studies reported loop features identified on the far-infrared,
these were either restricted to the Galactic midplane (Schwartz
\cite{Schwartz87}), performed for a special object type, e.g. Wolf-Rayet stars
(Marston \cite{Marston96}). Most of the prominent HI loops are conspicuous in the 
far-infrared, too, due to their dust content, 
which is closely related to HI (Boulanger \& Perault \cite{Boulanger88}). 
Many individual loops of various origin were first identified 
on FIR images. E.g. the North Celestial Pole Loop 
(Meyerdierks \cite{Meyerdierks91}) 
is an example of a high velocity cloud -- Galactic disc interaction. 
The \object{Cepheus Bubble} is a remarkable infrared loop
 (\cite{Kun87}) which is likely a combined result of multiple supernova 
 explosions and stellar winds in the Cep\,OB\,2 association.      

In a recent work Kiss et al. (2004, hereafter KMT04) presented the results of a
quest for far-infrared loop features in the \second Galactic Quadrant. They
catalogued 145 loops and investigated their morphological and physical
characteristics. This study was not restricted to the Galactic midplane, but
reached high Galactic latitudes and was able to give a
comprehensive view on the distribution of large scale intensity enhancements. 

In this present work -- as a continuation of the KMT04 study -- we extend the
quest to the \first, \third and \fourth Galactic Quadrants. These two works
together provide the {\it Catalogue of Far-InfraRed Loops in the Galaxy}. 

\section{Input data and data analysis}

We used the 60 and 100\,$\mu$m ISSA plates (IRAS Sky Survey Atlas,
Wheelock et al. \cite{Wheelock94}) in order to explore the distribution of 
dust emission. Reprocessed IRAS 100\,$\mu$m and reddening maps by Schlegel et
al. \cite{Schlegel98} were investigated as well. 
We searched for loop- or arc like features in these images in 
the 1$^\circ$\,$\leq$\,$D$\,$\leq$\,40$^\circ$ diameter range. 
The data reduction steps and derived parameters are the same as 
described in KMT04. For the detailed description of the data reduction 
and derived parameters we refer to KMT04 and Appendix~A (online only)
of this present paper. 

\section{Results}

In our all-sky survey 462 FIR loops have been identified.
We found 317 loops in the \first, \third~and \fourth~Galactic
Quadrants (presented in Appendix~B, online only), 
beyond the 145 loops in KMT04 (\second Galactic Quadrant). 
The electronic version of the catalogue with additional data products
can be found at: "http://kisag.konkoly.hu/CFIRLG".

The catalogue contains the name of the loop; the cental Galactic
coordinates, size and position angle of the fitted ellipse; and other 
parameters describing the appearance of the loop in FIR images. For
details, see Appendix~B in the electronic version of the paper or Sect.~3 in KMT04.

\noindent For a subsample of 43 loops of our catalogue 
(\first, \third and \fourth Galactic Quadrants) we were able to derive distances,
using the distances of associated objects. The selection criteria for distance
indicators were the following: \begin{itemize} \item[i)] objects related to
possible energy injection sources (O/B stars, SN-remnant, etc.)  had to be placed
in the interior of the loop; \item[ii)] density enhancements of the ISM
(molecular/dark clouds, molecular cores, etc.)  had to be in the interior or in
the loop wall; \item[iii)] there had to be at least two objects with similar
distances; \item[iv)] distances of individual objects had to agree within the
uncertainties. \end{itemize}  In many cases the distance of a loop has already
been determined by previous studies. In Appendix~C we present the distances and
a detailed list of distance estimators for these 43 loops. Fig.~\ref{dist76}
shows the distribution of estimated distances and the distances projected to the
Galactic plane.

In most cases, however, no definite distance could be derived for a specific
loop. In these cases we listed all possible associated objects (not taking
into account the distances of the individual objects), which appear on the loop
wall or in the interior. The list of these associated objects is presented 
in Appendix~D (online only, available at CDS: {\it http://...}).

\section{Discussion}  

\subsection{Sky distribution}

The Galactic longitude distribution of the GIRLs partly reflect the spiral
structure of the Galaxy (Figs.~\ref{arms} and \ref{H70}). There is a noticeable
increase in loop counts towards the major spiral arms i.e. the Local-, the inner
Carina-Sagittarius- and the outer Perseus Arms.

\begin{figure}
\centering  \includegraphics[width=6.5cm, angle=90]{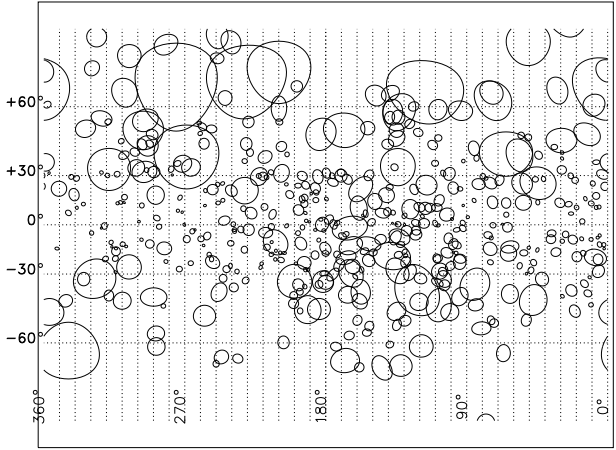} 
\caption{Distribution of GIRLs in the sky (Galactic coordinate system, Mercator projection), represented by the fitted ellipses. Note that Mercator
projection causes a size distortion at polar regions.}  
\label{sky}  
\end{figure}  

The distribution of GIRLs in the sky is expected to reflect the exponential disc
distribution of the ISM, accordingly regions closed to the Galactic plane should be the most
populated parts of sky in loops and this number should decrease nearly
exponentially by the increasing Galactic latitude. 

\begin{figure}[!h]  
\centering  
\includegraphics[width=8.0cm]{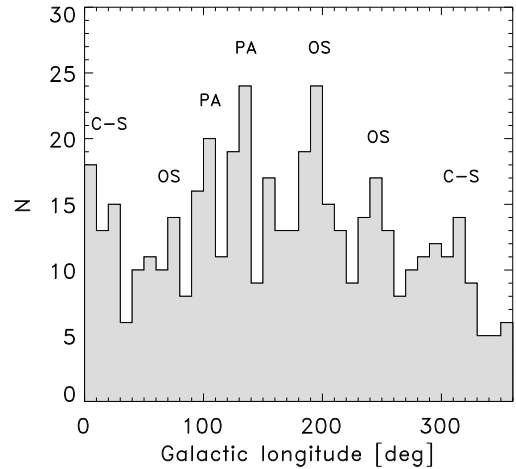}   
\caption{Distribution of loops in Galactic longitude. The peaks in loop-count reflect the 
higher number density in the direction of Galactic arms (OS: Orion Spur, PA: Perseus Arm, 
C-S: Carina-Sagittarius Arm). The binsize is 10\degr.} 
\label{arms}
\end{figure}  

The distribution of loop centres in Galactic latitude (Fig.~\ref{innerouter})  
-- which is expected to reflect the exponential disc distribution of the ISM --
shows relatively large counts at high $|b|$ values, especially for outward
locations. This cannot be explained by projection and distance effects only, in
agreement with that was found for the 2$^{nd}$ Galactic Quadrant (KMT04). To
explain the formation of loops at high Galactic latitudes one needs an efficient
process which does not necessarily have to take place close to the Galactic plane. 
This excludes SN-explosions and the stellar wind of massive stars as dominant effects. 
Although clouds infalling from the Galactic halo could create loops above the midplane, 
their infall rate is insufficiently small (\cite{Ehlerova96}).  

\begin{figure}[!h]   
\centering
\includegraphics[width=8cm]{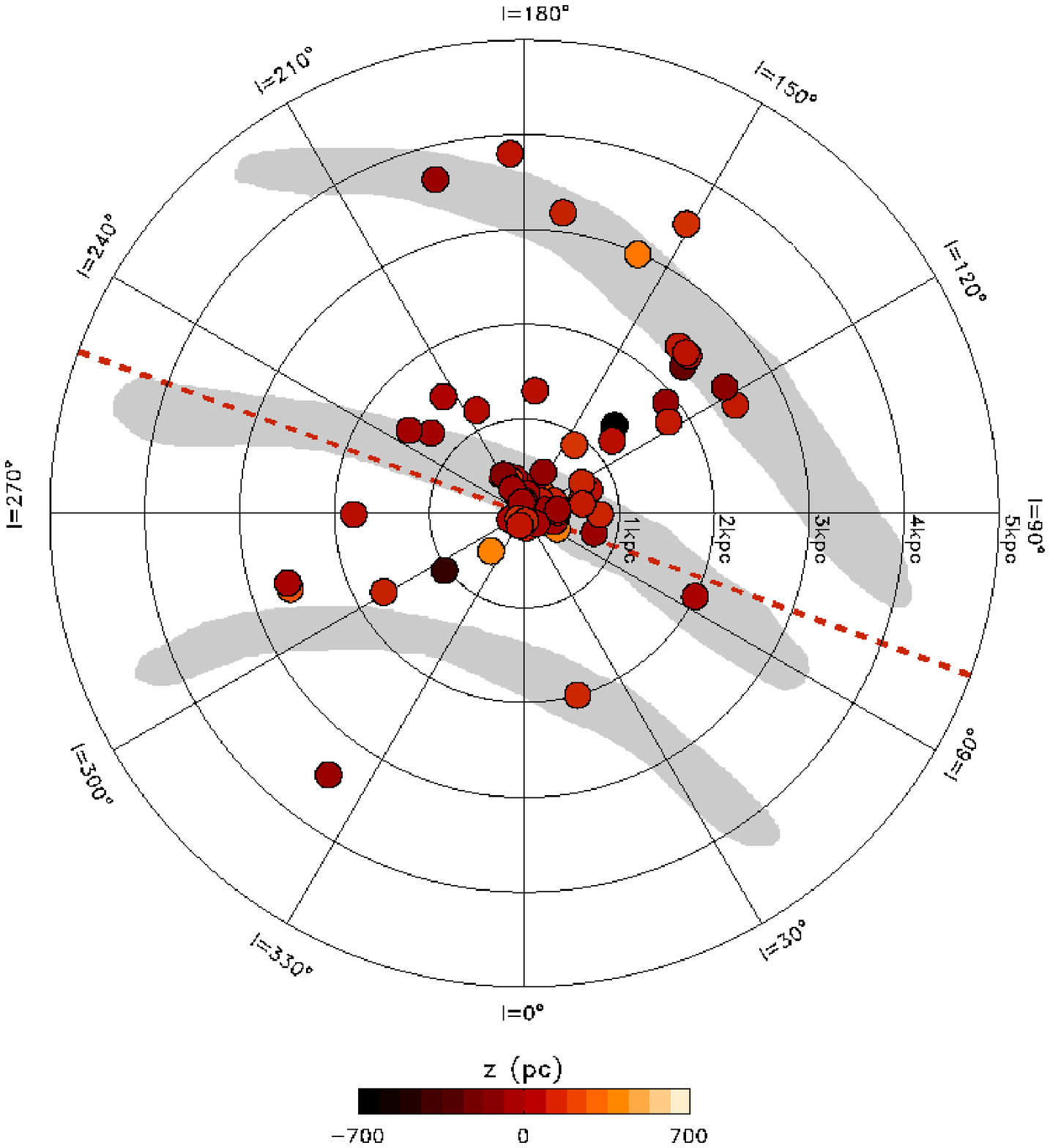}
\includegraphics[width=8cm]{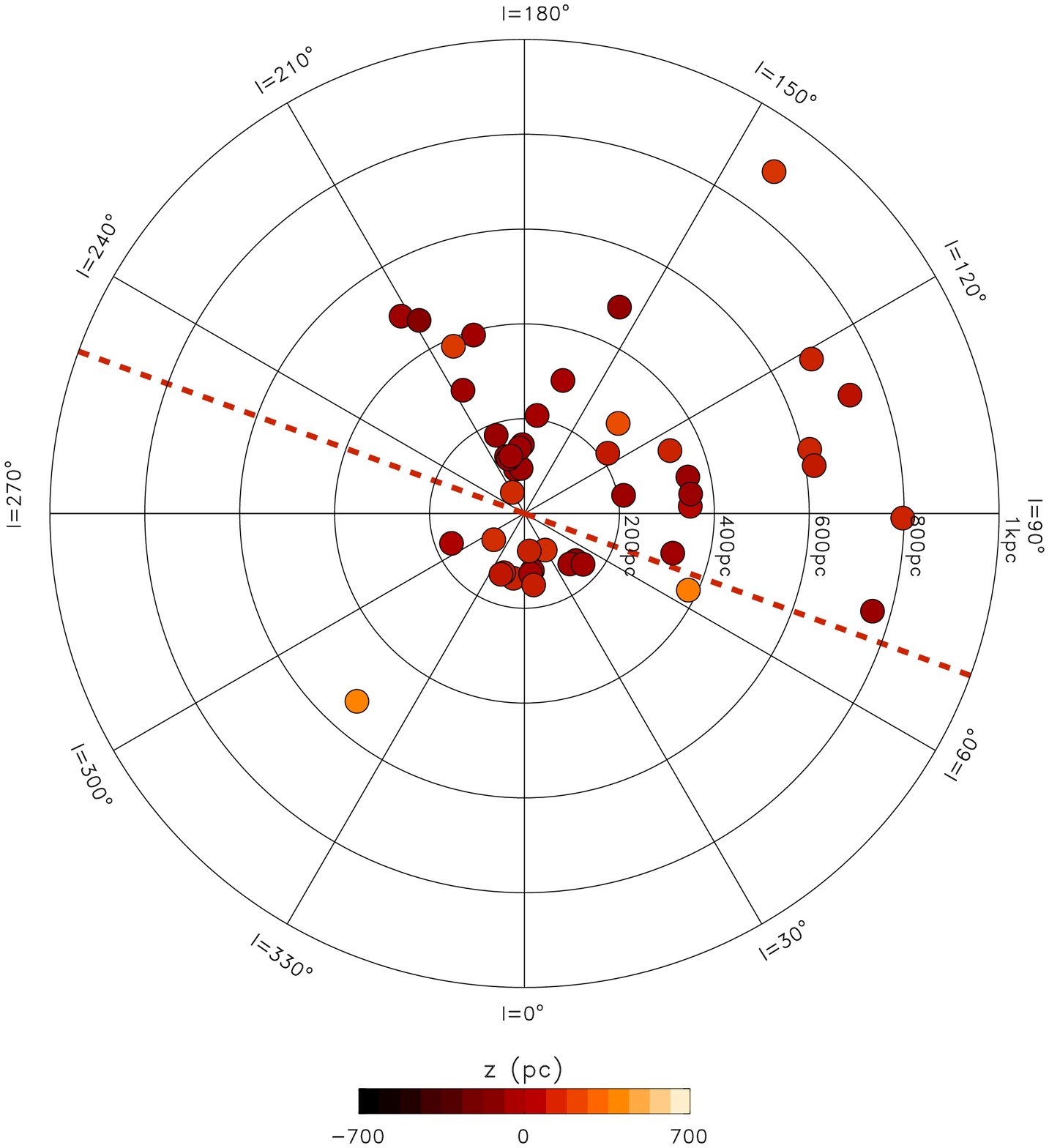}  
\caption{Galactic distribution of loops with known distances around the Solar System (located at the panel
centres). Top panel: location of infrared loops projected to the Galactic
plane, up to a distance of $d$\,=\,5\,kpc. Bottom panel: magnification of the central
region ($d$\,$<$\,1\,kpc). The distance from the Galactic midplane is indicated by
colors as given by the color bars. The approximate position of Spiral Arms (grey
areas) were adapted from \cite{Georgelin76}. The immediate \textit{outer}
surrounding of the Solar System was defined between $l$\,=\,70$\degr$--250$\degr$,
while the \textit{inner} Galaxy space lies on the other side of the
$l$\,=\,70$\degr$--250$\degr$ axis (dashed red line, see also Table~1). 
The majority of the loops are probably located in the Local Arm, as 
explained in Sect.~4.4).}   
\label{H70}  
\end{figure}  
\begin{figure}[!h]   
\centering
\includegraphics[width=8cm]{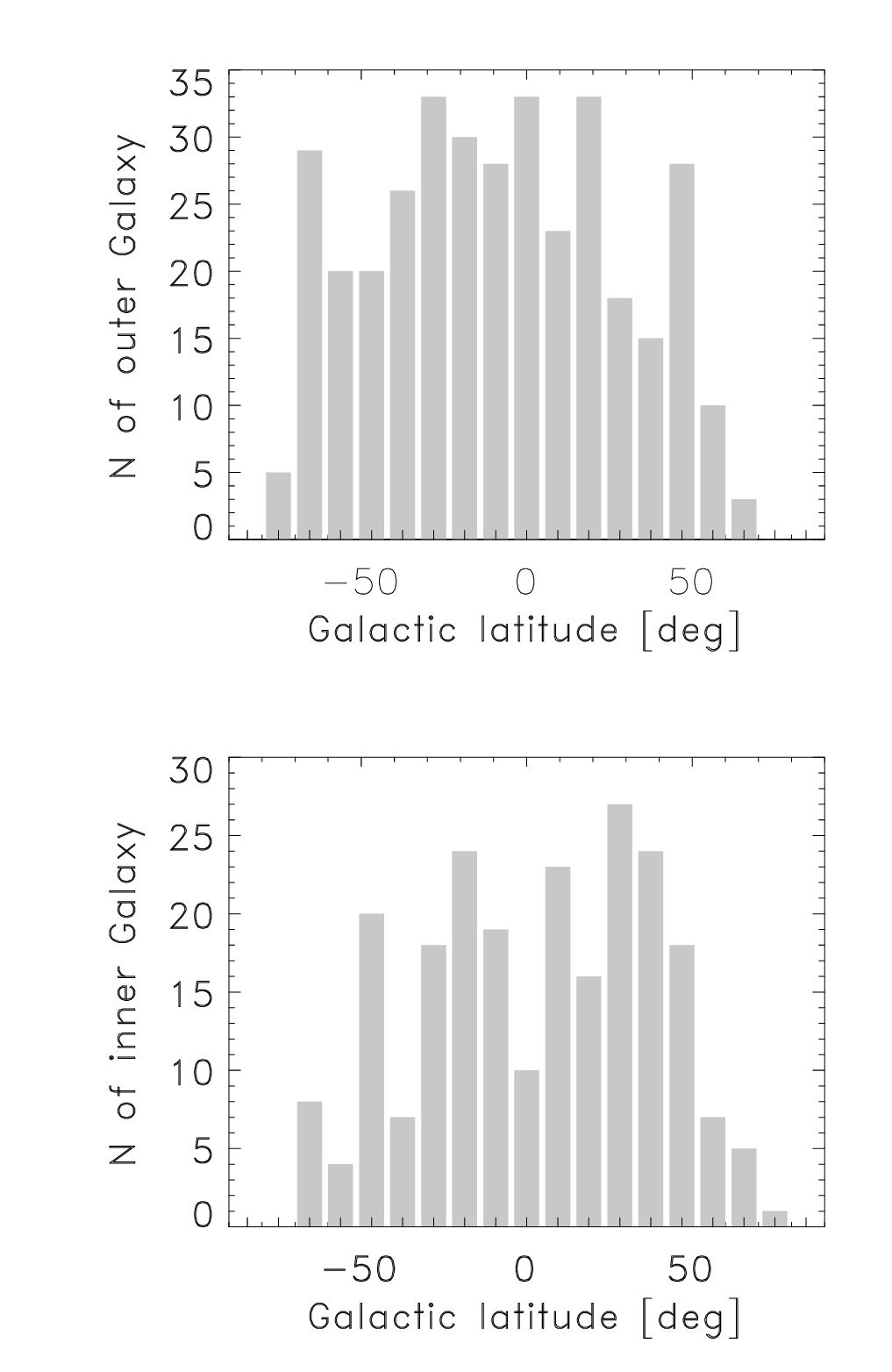} 
\caption{Galactic latitude distribution of loops in the outer  (B\,+\,D, top panel) and  inner (A\,+\,C,
bottom panel) Galaxy. See Table~1. The binsize is 10\degr.}   
\label{innerouter}   
\end{figure}  %

\subsection{Groups at larger scales}

The celestial distribution of the identified loops is rather complex  showing
structures with scales even larger than the average loop diameters
(Fig.~\ref{sky}). E.g. two remarkable features, forming 'chains' of loops can be
found around $l$\,$\approx$\,175$\degr$, $b$\,$\approx$\,+10$\degr$ and around
$l$\,$\approx$\,115$\degr$, $b$\,$\approx$\,--15$\degr$. This latter huge structure
is well seen as a fragmented arc in HI too, at velocities approximately --40
kms$^{-1}$ in the Leiden-Dwingeloo Survey (Hartmann \& Burton 1997).   

Some groups of loop centres coincide well with
the location of molecular complexes, including the Ophiucus Molecular Cloud
($l$\,$\approx$\,10$\degr$, $b$\,$\approx$\,+15$\degr$), the Ursa Maior Molecular
Cloud ($l$\,$\approx$\,135$\degr$, $b$\,$\approx$\,+55$\degr$), the Taurus molecular
complex ($l$\,$\approx$\,180$\degr$, $b$\,$\approx$\,+0$\degr$) and the
Orion-Monoceros molecular region ($l$\,$\approx$\,210$\degr$,
$b$\,$\approx$\,--15$\degr$).

Although it is likely, that loop structures are 
associated with large scale molecular material, no other 
obvious occurrences were found on these scales in our sample.

\subsection{Dominated by confusion?}

To independently study the immediate inner and outer surrounding of the Solar 
System, we arbitrarily split the Galactic latitude space into two parts, 
approximately along the direction of the Local Arm, i.e. the
$l$\,=\,70$\degr$--250$\degr$ axis.  The loop counts in the inner and outer part
show significant differences: there are almost twice as many loops in the outer
region than in the  inner one. One possibility to explain this fact is the
existence of  strong confusion in the direction of central (inward) Galactic
regions, as discussed below (for the definition of Galactic regions A, B, C, D see
Table~1). 

\begin{figure}[!!!h]   
\centering
\includegraphics[width=8cm]{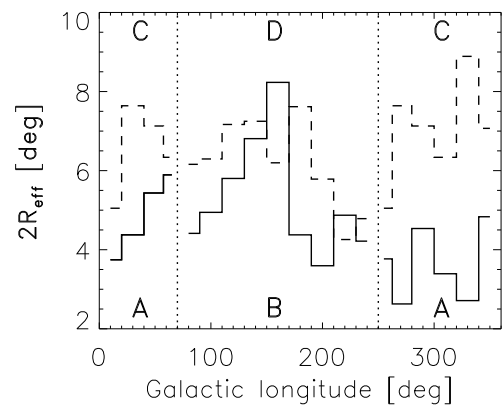}   
\caption{Average effective loop diameter vs. Galactic longitude in the inner (A\,+\,C) and outer (B\,+\,D)
Galaxy; solid and dashed lines correspond to $|b|$\,$\le$\,30\degr and
30\degr\,$\le$\,$|b|$\,$\le$\,60\degr, respectively. The binsize is 20\degr.}  
\label{glong-size} 
\end{figure}

\begin{figure}[!h]   
\centering 
\includegraphics[width=8cm]{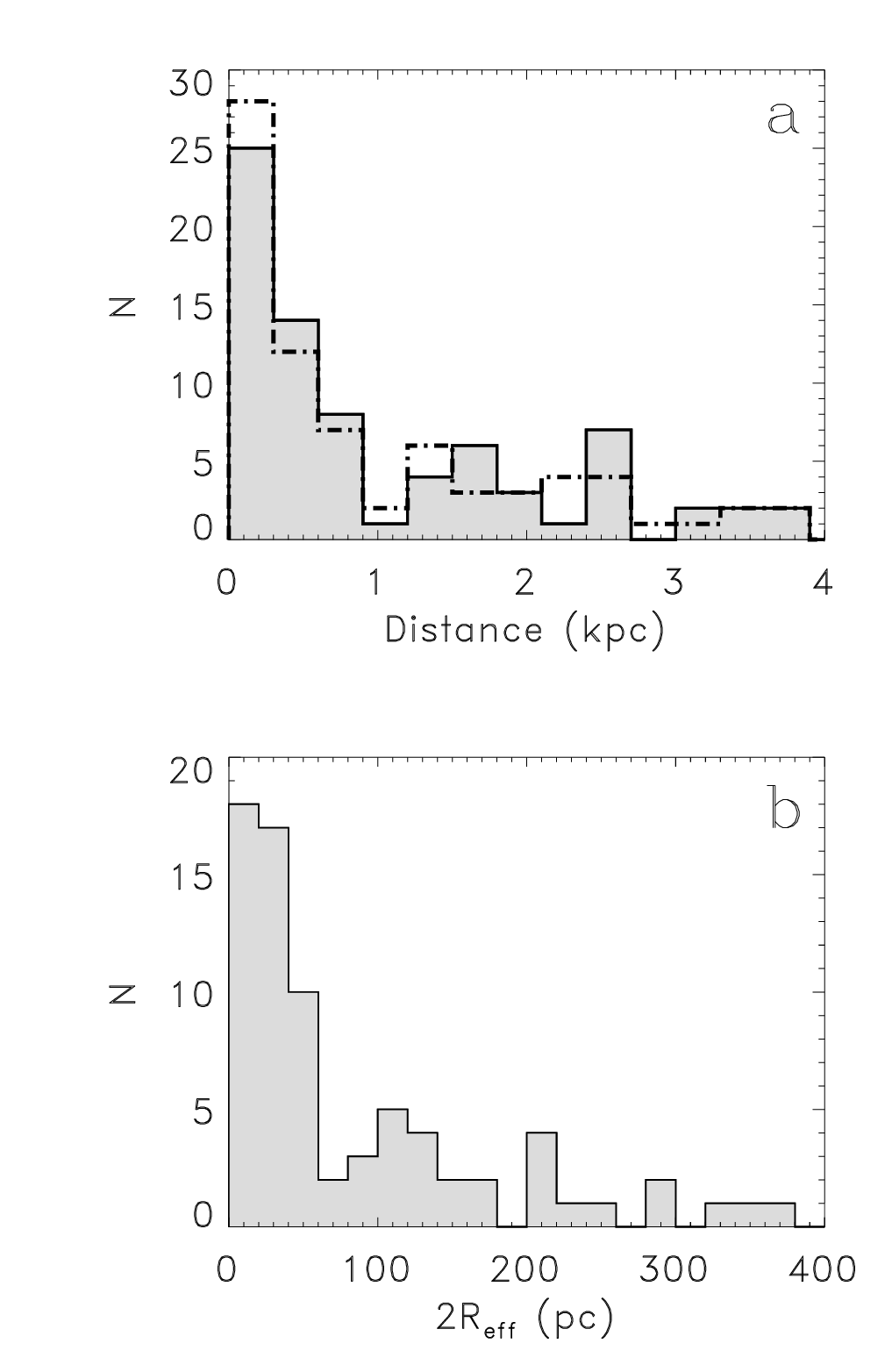}  
\caption{({\bf a}) Distribution of estimated distances for 73 loops. Dash-dotted line represents the 
distribution of distances \textit{projected} to the Galactic plane. ({\bf b}) Distribution of effective 
diameters in the same sample. The widths of the bins are 300\,pc and 20\,pc for (a, b), respectively.} 
 \label{dist76} 
  \end{figure}

The possibility of a confusion-effected distribution for the  2$^{nd}$ Galactic
Quadrant was already mentioned in KMT04.  In the present study -- due to the
extension to the whole sky --  this confusion  effect is more expressed and
manifests itself in two main ways :   
\begin{itemize}   
\item[i)] As presented in Fig.~\ref{innerouter}, we
see more loops in the 'outer' Galaxy than in the 'inner' regions for
$|b|\,<$\,30\degr (region~A),  although probabilities for the frequency of
high-pressure events (e.g. SN-rates) should be significantly higher in the inner
Galaxy.   
\item[ii)] The size distribution is different for regions A and B
(see Fig.~\ref{glong-size}):  in the inner Galaxy (A) the average apparent size
is smaller than in B (4\fdg1 and 5\fdg2, respectively).
This discrepancy is not observed  for higher Galactic latitudes (C and D).   
\end{itemize}

\noindent These two facts together indicate that  (a) either we miss large loops
at low $|b|$  in the inner Galaxy due to the strong IR background (small loops
are still visible since they are smaller than the characteristic scale of
molecular clouds) or (b) large loops are really  destroyed by frequent violent
events and are only visible at their early  evolutionary phases (i.e. at small
size). Unfortunately our present  FIR data cannot distinguish between these
possibilities. 

\begin{table}   
\begin{tabular}{|l|c|c|} \hline  
Region & Galactic Longitude & Galactic Latitude \\ \hline \hline  
A & $l$\,$<$\,70$\degr$ or $l$\,$>$\,250$\degr$ & $|b|$\,$<$\,30$\degr$ \\  
& inner & low \\ \hline   
B & 70$\degr$\,$<$\,$l$\,$<$\,250$\degr$ & $|b|$\,$<$\,30$\degr$ \\  & outer & low \\
\hline   
C & $l$\,$<$\,70$\degr$ or $l$\,$>$\,250$\degr$ & 30$\degr$\,$\le$\,$|b|$\,$\le$\,60$\degr$ \\ 
 & inner & high \\ \hline  
 D & 70$\degr$\,$<$\,$l$\,$<$\,250$\degr$ &
30$\degr$\,$\le$\,$|b|$\,$\le$\,60$\degr$ \\   & outer & high \\  \hline 
\end{tabular}   
\caption[]{Definition of four Galactic regions. Regions
have been defined to help the comparison of loop characteristics at different
parts of the sky. }  
 \label{Subregionstable}   
 \end{table}

\subsection{The large scale distribution of ISM in the Galaxy}

%
\begin{figure}[!!h]   
\centering 
\includegraphics[width=8cm]{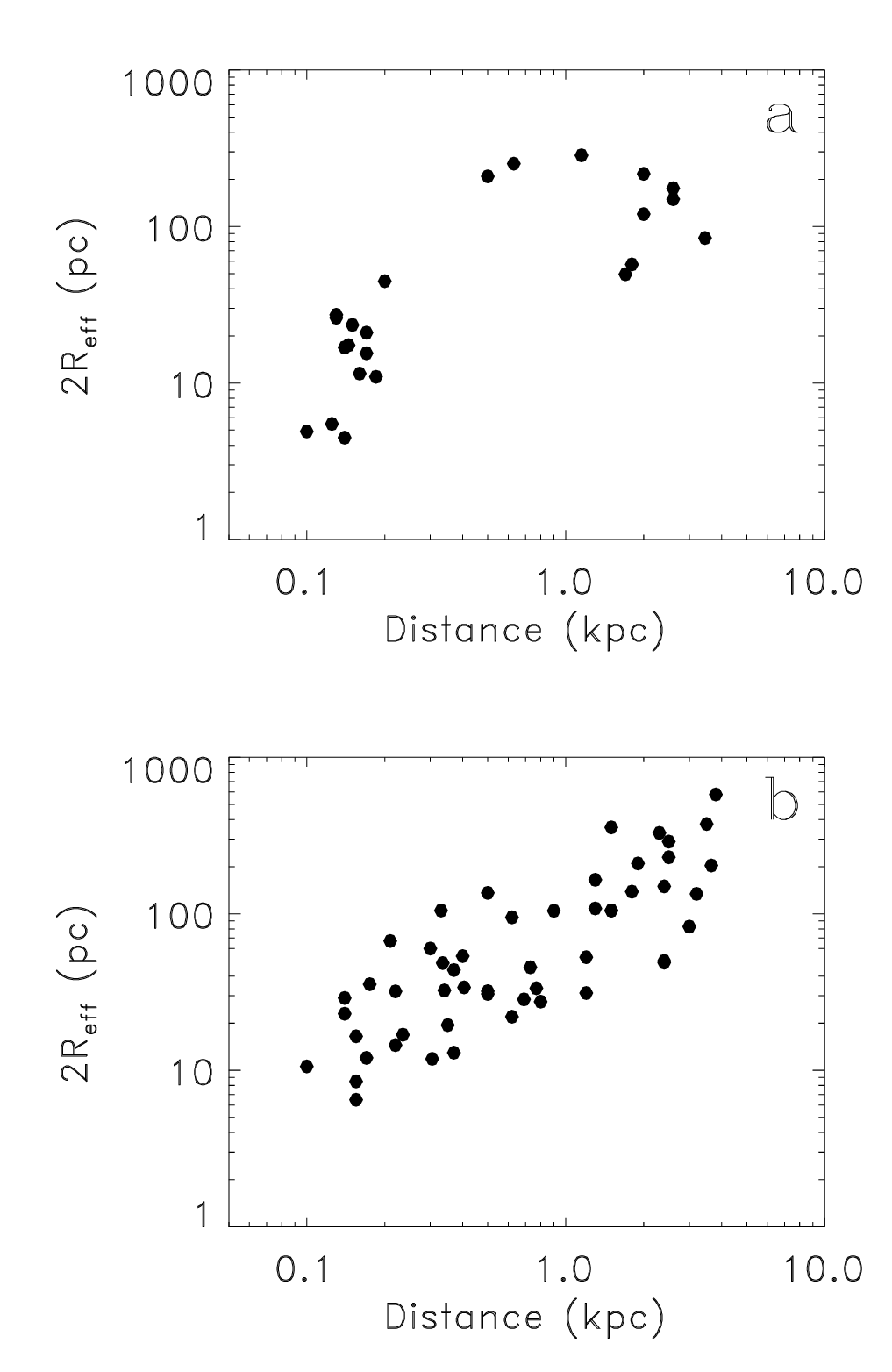}  
\caption{Correlation of distance and effective size in the "inner"  
({\bf a}) and in the "outer" Galaxy
({\bf b}), based on the sample of 73 loops with known distances.}  
\label{dist_vs_diam}   
\end{figure}  

\begin{figure}[!h]   
\centering
\includegraphics[width=8cm]{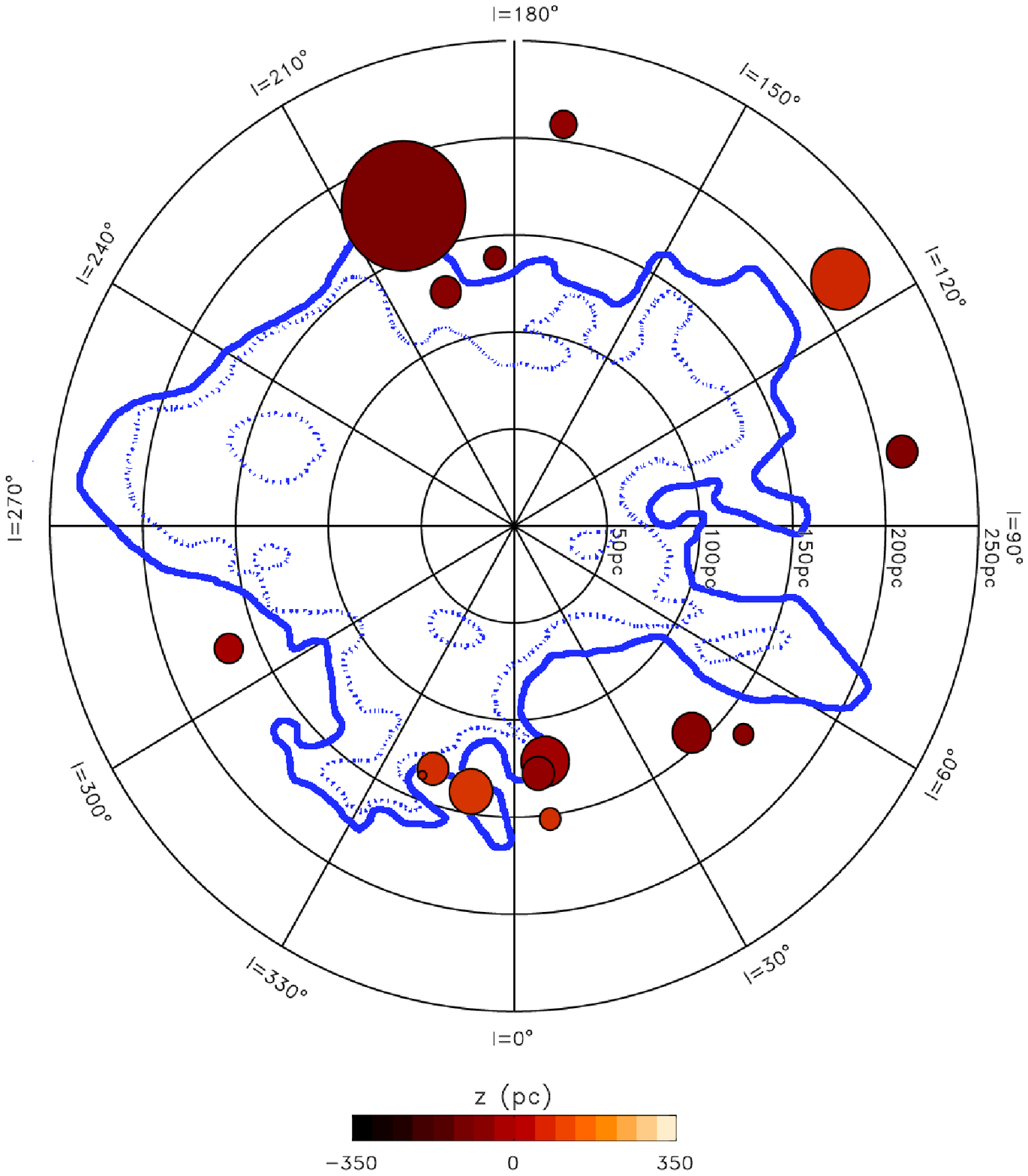}
\caption{Location of known distance loops in regions A and B in the close vicinity of the Solar System. The size of the
filled circles indicates the real effective size of the loops. Dashed and solid blue curves correspond to the
W(D2)\,=\,20\,m{\AA} and W(D2)\,=\,50\,m{\AA} contours of equivalent width of the NaI doublet line (see \cite{Lallement} 2003).}    
\label{lallement}  
\end{figure}

In Appendix~C we presented a subsample of 43 loops of our catalogue.  For these
loops we were able to derive distances using the distances of associated objects.
Together with the 30-loop sample in appendix~B in KMT04 we have a sample 
of 73 loops with known distances. We used this sample to characterize the 
large-scale distribution of the ISM in the vicinity of the Solar System. 

\subsubsection{Size distribution of cavities} 

In Fig.~\ref{dist_vs_diam} we 
present the differences in distance distributions in the inward and outward Galactic regions.
In the inward region A most of the loops are  squeezed into a small area around the
Solar System, within a distance of $\sim$0.2\,kpc (see Fig.~7). 
The rest of the loops do not show a concentration at a specific distance but are rather randomly distributed 
in the interval \lele{0.4\,kpc}{$d$}{4\,kpc}. 
Note, that the 2$^{nd}$ Galactic Quadrant has a major contribution to
outward regions, therefore this distribution is quite similar to the one found in KMT04.

We checked the distances of the individual loops (those with known
distances in regions A and B) 
and the distance of the wall of the \object{Local Bubble} in the direction of 
a specific loop using the maps and data based on the measurements of the
NaI D-line doublet by \cite{Lallement} (2003). The reassuring 
result of this comparison was, that all of these loops were in or behind the wall, i.e. behind the
W(D2)\,=\,20\,m{\AA} contour of equivalent width (see Fig.~\ref{lallement}).

\begin{figure}[!h]  
\centering 
\includegraphics[width=8cm]{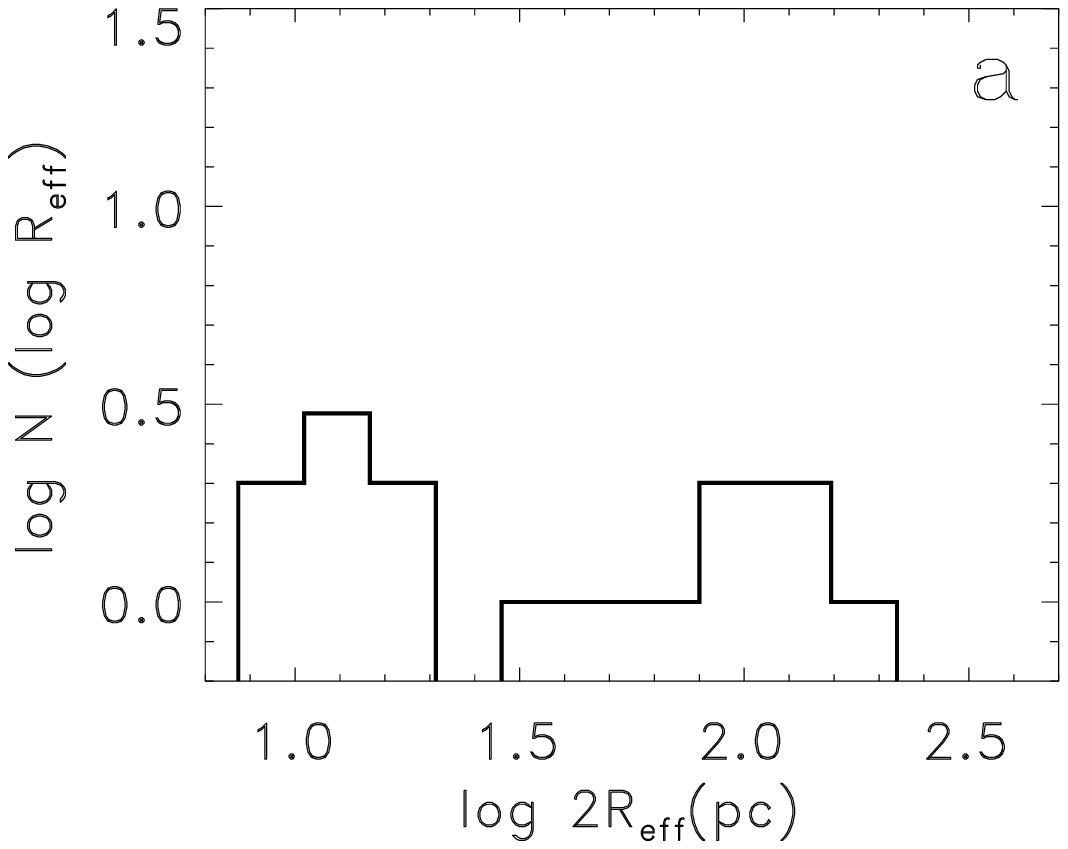} 
\includegraphics[width=8cm]{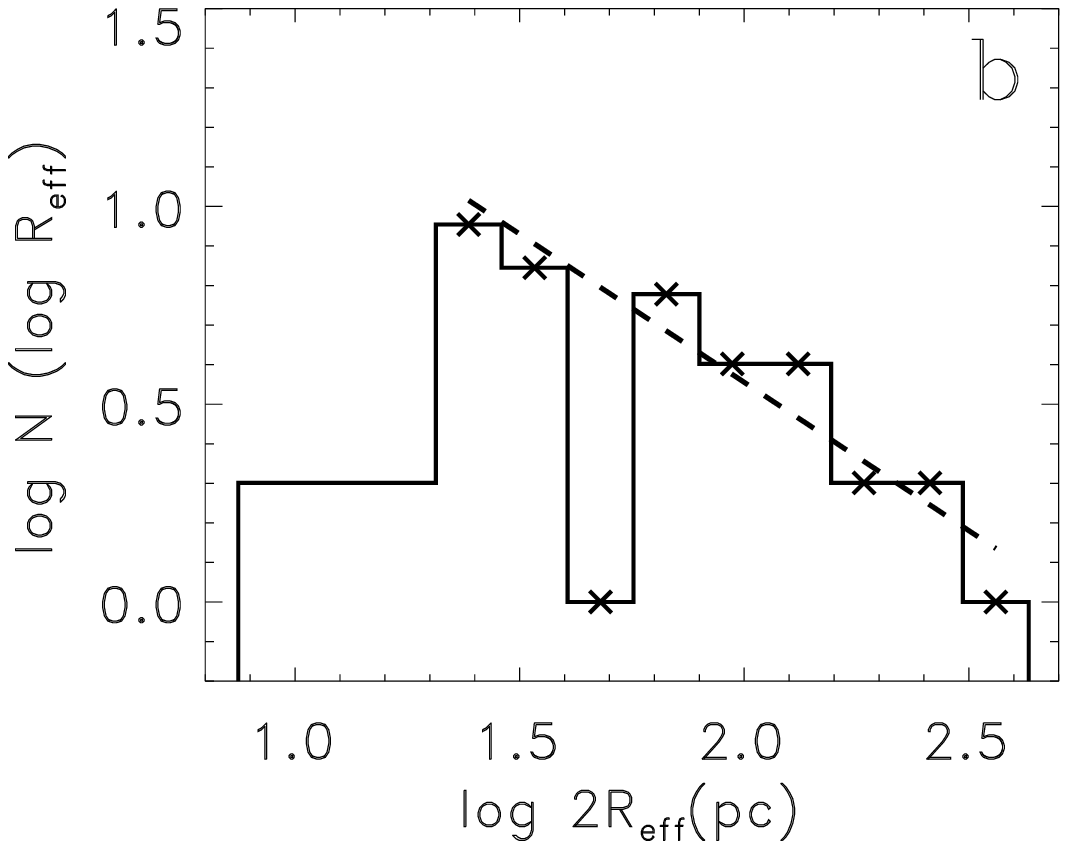}  
\caption{Distribution of effective diameters for known-distance loops on a logarithmic scale in the
"inner" ({\bf a}) and "outer" ({\bf b}) Galaxy. The relation of log2$R_{\mathrm{eff}}$-log$N$ was fitted for
1.4\,$\leq$\,log(2$R_{\mathrm{eff}}$)\,$\leq$\,2.6. The points used for the fit are marked by
asterisks (see Appendix~B).}   
\label{effdiam76}   
\end{figure}

The general structure of galaxies is often characterized by the distribution 
of bubbles and superbubbles, as was done for nearby galaxies  
by Oey \& Clarke \cite{Oey97} and Kim et al. (\cite{Kim03}). 
The generally used parameter $\beta$, the power-law index of the mechanical
luminosity function, is closely related to the power law index $s$
of the logarithmic size distribution: $s$\,=\,2--2$\beta$ 
(see the model by Oey \& Clarke \cite{Oey97}).
Using this relationship we derived a power law 
index of $\beta=1.37\pm0.17$ for our outer sample (region B) for the interval
1.4\,$\le$\,log(2$R_{\mathrm{eff}}$)\,$\le$\,2.6, which is similar to the value of the
2$^{nd}$ Galactic Quadrant only ($\beta=1.24\pm0.30$, KMT04) 
and is in good agreement with $\beta=1.6\pm0.3$ 
obtained by Ehlerov\'a \& Palou\v{s} (2005). Our value is 
lower than that of the other investigated galaxies (see Fig.~\ref{effdiam76}b). 
The double-peaked distribution observed in KMT04 is less expressed in this larger 
sample and is closer to a power-law. Due to the insufficient sample no proper 
$\beta$ could be derived for the inner Galaxy (region A), however, 
the distribution is very shallow (Fig.~9a).    

We have no information on the distances of loops at high Galactic 
latitudes due to the lack of proper distance indicators. This prevented 
us from deriving proper $\beta$ values for high Galactic latitudes 
(C and D) as we did for A and B. However, the distribution of apparent size  
of the known-distance sample and that of the full high Galactic latitude 
sample are remarkably  different. This
indicates different real size distributions behind the apparent ones, 
and the presence of a different process governing the structure. 

Due to the lack of distance information only the apparent size 
distribution can be used as diagnostic tool to test the structure of
the high Galactic latitude ISM. In a simple model we tested the 
apparent size distribution with different $\beta$ values. 
The model included a real size distribution with a prescribed $\beta$
value (so that the slope of the log($R$) vs. log($N$[log($R$)]) distribution
is $s$\,=\,2--2$\beta$) and a certain distribution of cavity centres. 
We derived the
apparent size distribution from this size/spatial location distribution.
The apparent size distribution was then fitted by a power law (as in the case
of the real size distribution), with a power law index of $s_{app}$, 
in the interval \lele{1\degr}{2$R_{\mathrm{app}}$}{40\degr}, i.e. in our
investigated apparent diameter (2$R_{\mathrm{app}}$) range. 

The main features and results of this simple model were the following: 

\begin{itemize}

\item Assuming a random, homogeneous distribution of cavity centres, with
a minimal distance of d$_{\mathrm{min}}$\,=\,0\,pc, the final $s_{\mathrm{app}}$
has little dependence on the absolute value of the maximal distance d$_{\mathrm{max}}$,
as long as the apparent sizes are in the fitted range. 
Realistic estimates of the maximal distances
are from $\sim$250\,pc (scale height of the Galactic disc; Nakanishi \& Sofue, 2003) 
to a few kpc (the maximal visibility distance), 
depending on the Galactic region we try to model.

\item The assumption of d$_{\mathrm{min}}$\,=\,0 is not realistic, 
since the Solar System is located inside the Local Bubble, and the loops must be
behind its wall, as it is the case for the objects in our known-distance sample. 
We choose a minimal distance of d$_{\mathrm{min}}$\,=\,130\,pc
(a characteristic distance of the wall of the Local Bubble, see Lallement et al. 2003), 
while keeping the random, homogeneous distribution behind. In this configuration
the apparent size distribution is clearly altered, compared to the 
d$_{\mathrm{min}}$\,=\,0 cases. The apparent size distribution is shallower, and
the larger the maximal distance, the closer the 
$s_{\mathrm{app}}$ values with d$_{\mathrm{min}}$\,$\ne$\,0 to the 
ones with d$_{\mathrm{min}}$\,=\,0. Although the difference is noticeable, 
$s_{\mathrm{app}}$ has a stronger dependence on $\beta$ then on d$_{\mathrm{min}}$
(not considering very extreme conditions). 

\item We have checked with a Kolmogorov-Smirnov (KS) test, whether the
apparent size distribution of high la\-titude loops in the northern and
southern hemispheres may originate from the same distribution. The KS
test resulted in a significance level of $\sim$50\%, showing that the two
distributions are not really different. The $s_{\mathrm{app,high}}$ values derived
independently for the two hemispheres resulted in very similar values,
therefore we used the data of the two hemispheres together in the
following. There is a noticeable difference in the total counts (110 and
78 at the northern and southern hemispheres, respectively), but this
does not affect the $s_{\mathrm{app,high}}$ values.

\item Assuming $\beta$\,=\,1.37 (that of the real size distribution in the
known-distance sample), the apparent size distribution has 
$s_{\mathrm{app}}$\,=\,--0.7 for the $d_{\mathrm{min}}$\,=\,0, and 
\lele{--0.7}{$s_{\mathrm{app}}$}{--0.5} for the d$_{\mathrm{min}}$\,=\,130\,pc
case, choosing the maximal distance in the range 
\lele{250\,pc}{d$_{\mathrm{max}}$}{1\,kpc}. This shows a good agreement 
with the power law index of the apparent size distribution of the known-distance
sample, $s_{\mathrm{app,known}}$\,=--0.67$\pm$0.12. 

\item The known-distance sample is characteristic for low Galactic latitudes 
(b\,$\le$\,30\degr) only. The power law index of $s_{\mathrm{app,high}}$\,=--1.78$\pm$0.21 
is quite different from $s_{\mathrm{app,known}}$. The likely difference is a 
different real size distribution, i.e. a different $\beta$.
We tested, whether a realistic $\beta$ can preproduce
this value: $\beta$\,=\,2.15 ($s$\,=\,--2.3, the canonical value
for a fractal structure created by supersonic turbulence) 
resulted in $s_{\mathrm{app}}$\,=\,--1.8 for the d$_{\mathrm{min}}$\,=\,0
and \lele{--2.2}{$s_{\mathrm{app}}$}{--1.8} for the d$_{\mathrm{min}}$\,=\,130\,pc
case. These values are indeed very similar to $s_{\mathrm{app,high}}$.

\item We checked whether a size distribution with $\beta=1.37$ 
may be responsible for an apparent size distribution of 
$s_{\mathrm{app}}\,=-1.78$, assuming a specific spatial distribution 
of cavity centres. 
To reproduce the required values of both parameters 
at the same time, the spatial distribution has to be so, 
that the loops are highly concentrated at the outer edge 
of the investigated volume and practically no loops are found at 
small distances. This is an unlikely configuration even taking to 
account that the Solar System is located inside the Local Bubble.

\end{itemize}

As we saw, various $\beta$ values lead to significantly 
different apparent size distributions, while changing the minimal and
maximal distances in a realistic interval alters the size distribution
less. Therefore it is not very likely, that very different 
$\beta$ values would lead to the same $s_{app}$ using 
different spatial distributions, in the framework of our model. 
We have to note, that among other limitation like the unknown real spatial 
distribution, our simple model do not take into account any 
observational biases, which are especially important at small 
apparent sizes. 

However, our simple model clearly outlines the difference in the 
{\it real} size distributions at low and high Galactic latitudes.
A probable explanation for this difference is that 
the structure is governed by different
physical processes. Our high latitude apparent size distribution
favours a $\beta$ value similar to that of supersonic turbulence
(Elmegreen, 1999), which has been long considered as the most likely
process forming the structure of the 
interstellar medium further from the Galactic midplane. 

\subsubsection{Volume filling factor}

The hot gas volume filling factor $f$ is predicted to change with the
Galactocentric distance (Ferriere \cite{Ferriere98} and 
Gazol-Pati\~no \& Passot
\cite{Gazol-Patino99}): it is 20...30\% 
for the inner Galaxy and decreases
significantly for larger Galactocentric distances. 
For low Galactic latitudes ($b$\,$\le$\,30\degr) -- where the structure 
is mostly formed by violent events -- one can assume, that the distribution 
of hot gas is very near to the distribution of holes/cavities and therefore 
can be characterized by their distribution.  
Applying the same procedure as in KMT04 (see Sect~4 in KMT04, paragraph 
"Volume filling factor") for the whole loop sample in the inward and 
outward regions separately for $b$\,$\le$\,30\degr~ (regions A and B), 
we obtained $f_{\mathrm{in}}$\,$\approx$\,30\% and 
$f_{\mathrm{out}}$\,$\approx$\,5\% for 
A and B, respectively. 
This latter value is very similar to the one presented in KMT04 
for the 2$^{nd}$ Galactic Quadrant (4.6\%\,$\le$\,$f_{\mathrm{KMT04}}$\,=\,6.4\%). 
The size distribution in the inward 
region A almost completely misses the large
loops which might be due to confusion. 
However, these may not exist at all,  since large
structures are regularly destroyed by violent events. The $f_{\mathrm{in}}$ value
we found is in the order of $\sim$20\%, as predicted by Ferri\'ere
(\cite{Ferriere98}) and Gazol-Pati\~no \& Passot (\cite{Gazol-Patino99}). Our
$f_{\mathrm{out}}$ value is very similar to $f_{\mathrm{HI}}$\,$\approx$\,5\% found by Ehlerov\'a
and Palou\v{s} (\cite{Ehlerova05}) through an automated identification of HI
shells in the 2$^{nd}$ Galactic Quadrant.

\section{Summary} 

We performed a survey of loop/arc-like intensity 
enhancements in the diffuse far-infrared emission
in the Galactic longitude intervals 
0\degr\,$\le$\,$l$\,$\le$\,90\degr ~and 
180\degr\,$\le$\,$l$\,$\le$\,360\degr.   
Merged with the results of the second Galactic Quadrant (KMT04),
we altogether identified and catalogued
462 of these features. 

The {\it Catalogue of Far-Infrared Loops in the Galaxy}
contains the basic physical properties and a list of associated 
objects as well.
The electronic version of the catalogue is available at: {\it
http://kisag.konkoly.hu/CFIRLG/}.   
We also gave distance estimates for 73 loops based on the distance
of associated objects. 
Our database  provides a great opportunity to study the large scale 
structure of the ISM in the Galactic neighbourhood of the Sun.  
We determined
observational estimates for the hot gas volume filling factor of the the inner
($f_{\mathrm{in}}$) and outer ($f_{\mathrm{out}}$) Galactic environment of the Solar System and
obtained 
$f_{\mathrm{in}}$\,$\approx$\,30\% and 
$f_{\mathrm{out}}$\,$\approx$\,5\%. These values are in good
agreement with both theoretical models and values derived from measurements in the radio domain.  
A study of the cavity size distribution has shown that the structure
of the ISM is different at low and high Galactic latitudes. While 
at low latitudes it can be explained by high pressure events,
the apparent size distribution of holes favours a fractal structure
at high latitudes, similar to what can be created by supersonic 
turbulence.

\begin{acknowledgements} We are grateful to L.G. Bal\'azs and A. P\'al for their
kind help in the data analysis. This work was partly supported by the grants
T043773 and K62304 of the Hungarian Scientific Research Fund (OTKA). We are indebted to our 
referee, Dr. So\v{n}a Ehlerov\'a, for her useful comments, which helped us to improve the paper.
\end{acknowledgements}


\newpage

\appendix
\section{Detailed description of data reduction steps and main parameters}

\begin{itemize}
\item {\bf Identification of the loop structure:}
We investigated the 60 and 100\,$\mu$m ISSA plates 
(IRAS Sky Survey Atlas, Wheelock et al. 1994) to explore the distribution of dust 
emission. We created composite images of the 
$12\fdg5\times12\fdg5$ 
sized individual ISSA plates using the "geom" and "mosaic" procedures of the IPAC-Skyview 
package\footnote{http://www.ipac.caltech.edu/Skyview/}, 
both at 60 and 100\,$\mu$m. These images were built up typically 
from 10...15 ISSA plates, reaching a size of 
${\sim}\,40{\hbox{$^\circ$}}\times40{\hbox{$^\circ$}}$. 
Loop-like intensity enhancements were searched for by eye on the 
100\,$\mu$m mosaic maps. Loops by our definition must show an excess 
FIR intensity confined to an arc-like feature, at least 60\% of a 
complete ellipse-shaped ring. A loop may consist of a set of bright, 
more or less isolated, extended spots, or may be a diffuse ring or part of a ring. 
The size of the mosaic image limits the maximal size of the objects found. On the other 
hand, due to the relatively large size of the investigated regions, loop-like intensity 
enhancements with a size of $\le $1$^\circ$ were not searched for. The original ISSA 
$I_{\mathrm{60}}^{\mathrm{ISSA}}$ and $I_{\mathrm{100}}^{\mathrm{ISSA}}$ surface brightness 
values were transformed to the COBE/DIRBE photometric system, using the conversion 
coefficients provided by Wheelock et al. (1994):
\begin{itemize}
\item $I_{\mathrm{60}} = 0.87\times I_{\mathrm{60}}^{\mathrm{ISSA}} + 0.13~\rm MJysr^{-1}$
\item $I_{\mathrm{100}} = 0.72\times I_{\mathrm{100}}^{\mathrm{ISSA}} - 1.47~\rm MJysr^{-1}$. 
\end{itemize}
Dust IR emission maps by Schlegel et al. (1998) (SFD) were investigated to derive 
parameters describing our loop features (see Sect. 2.2). The main differences of 
the SFD 100\,$\mu$m map with the ISSA maps are the following: 
(1) Fourier-destriping was applied; 
(2) asteroids and non-Gaussian noise were removed; 
(3) IRAS and DIRBE 100\,$\mu$m maps were combined, preserving the DIRBE zero 
point and calibration; 
(4) stars and galaxies were removed. 

We analysed the radial surface brightness profiles of the loops on the 
SFD 100\,$\mu$m map in order to check the effect of the removal of the sources 
mentioned above. We also used the SFD E(B-V) maps derived from the dust column density 
maps. In the case of these maps the colour temperature was derived from the DIRBE 100 
and 240\,$\mu$m maps, and a temperature-corrected map was used to convert the 
100\,$\mu$m cirrus map to a map proportional to dust column density (see SFD).

\item {\bf Ellipse fitting:}
The shape of a loop candidate was approximated by an ellipse, which was then fitted 
using a 2D least-squares fit method. 
The fitted ellipse is defined with the central (Galactic) coordinates, the minor 
and major semi-axis of the ellipse, and the position angle of the major axis 
to the circle of Galactic latitude at the centre of the ellipse. 
This latter was defined to be "+'' from East to North (or counter-clockwise).

\item {\bf Intensity profile:}
For all of our loops we extracted radially averaged surface brightness profiles, 
extending to a distance of twice the major (and minor) axis of the fitted ellipse, 
using 40 concentric ellipsoidal rings. These surface-brightness profiles 
(ISSA 100 and 60\,$\mu$m, SFD 100\,$\mu$m and SFD reddening maps) were 
used in the following to determine the basic parameters of the FIR emission in the loop; 
two examples are presented in Fig.~A.1.

\begin{figure}[!h]   
\centering
\includegraphics[width=8cm]{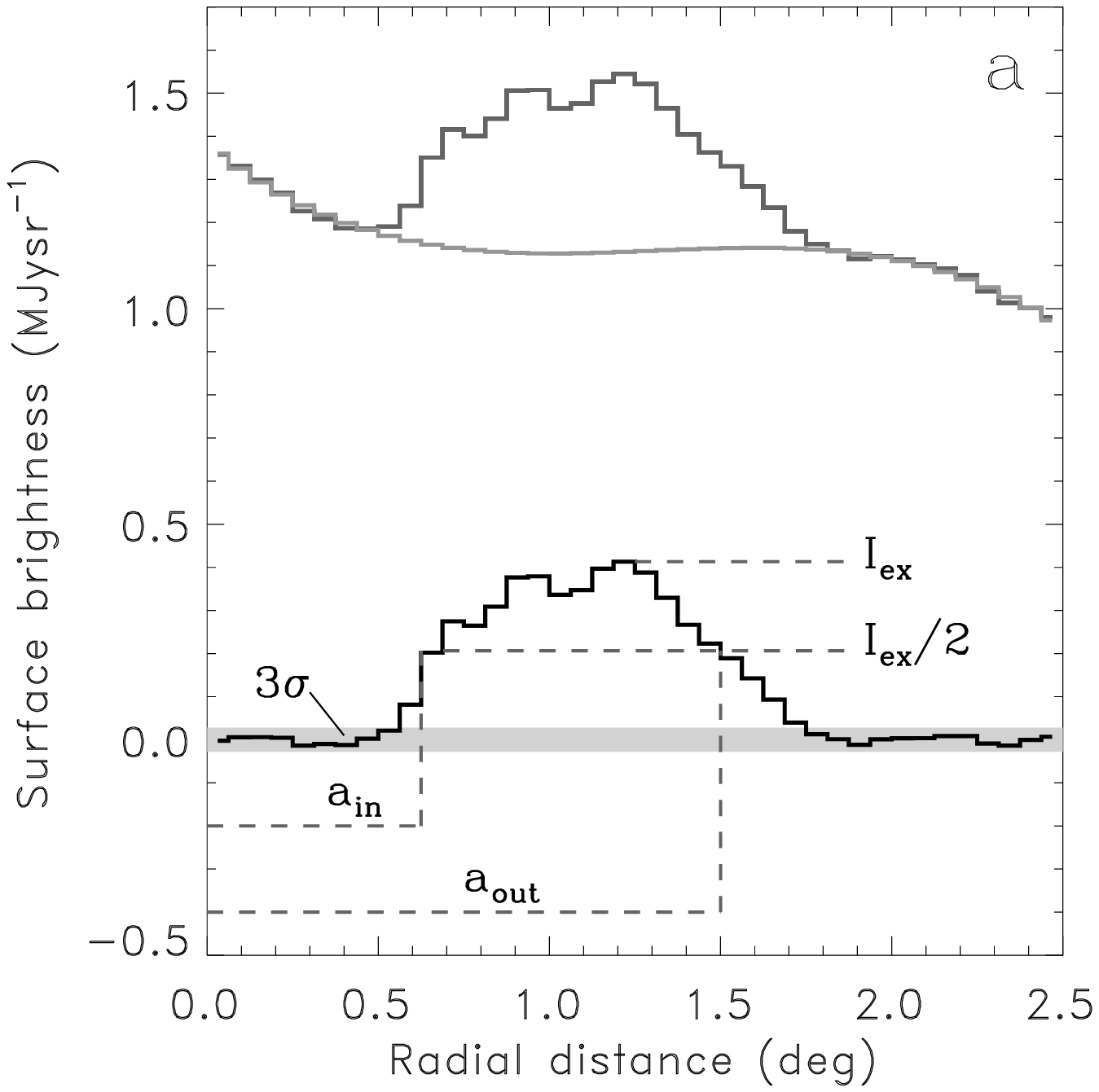}
\includegraphics[width=8cm]{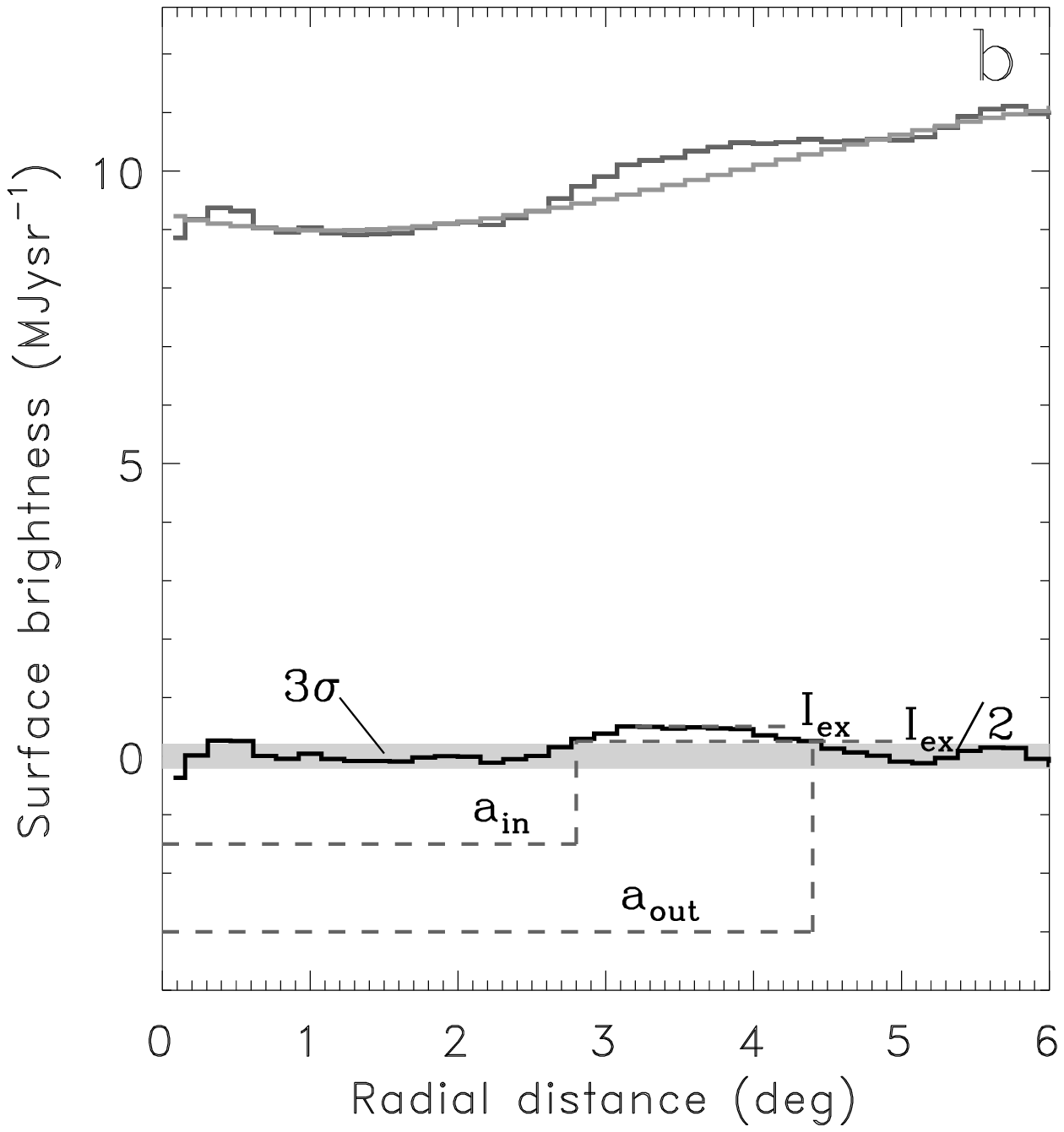}  
\caption{Example of a ({\bf a}) high significance loop 
(G077-77, $\Psi_{\mathrm{I100}}$\,=\,38.5) and a ({\bf b}) 
low significance loop (G019-19, $\Psi_{\mathrm{I100}}$\,=\,3.7). 
Besides low surface brightness enhancements, bright spot-like 
objects and point sources in the neighbourhood of the loop may 
also result in poor significance values as was demonstrated in KMT04. 
Figures show the intensity profile before (top) and after 
(bottom) background removal (grey solid line) with the main derived 
parameters ($I_{\mathrm{ex}}$, $\sigma_{\mathrm{ex}}$, $a_{\mathrm{in}}$, 
$a_{\mathrm{out}}$).}   
\label{loopcorr}  
\end{figure}  

\item {\bf Parameters of the loop wall:}
A local background was determined using the "non-loop'' points in the 
radial surface brightness profiles, fitting a 3rd order polynomial, as 
demonstrated in Fig.~A.1. 
This appropriate background was then removed from each surface brightness profile point. 
The intensity excess $I_{\mathrm{ex}}$ 
was derived as the maximum value of this background-removed profile. 
Inner and outer edges of the loop wall along the major axis 
$a_{\mathrm{in}}$ and $a_{\mathrm{out}}$, respectively, are 
defined as the radial distance at the full width at $I_{\mathrm{ex}}/2$, the half power of the 
background-removed intensity profile. 
We distinguish three regions for a specific loop: (1) loop interior 
($a\,<\,a_{\mathrm{in}}$); (2) loop wall 
($a_{\mathrm{in}}\,\le\,a\,\le\,a_{\mathrm{out}}$); 
(3) outer region ( $a\,>\,a_{\mathrm{out}}$). 
The relative width of the wall of the fitted ellipse is defined as 
$W$\,=\,$1-a_{\mathrm{in}}/a_{\mathrm{out}}$.

\item {\bf Significance:}
We calculated the standard deviation of the background-removed surface 
brightness, $\sigma_{ex}$, in the 'non-loop' positions (a\,$\le$\,a$_{in}$ and 
a\,$\ge$\,a$_{aout}$), and defined the significance of the loop as 
$\Psi = I_{\mathrm{ex}}/ \sigma_{\mathrm{ex}}$.  
We derived significance parameters on 100 and 60\,$\mu$m ISSA maps and on the 
SFD 100\,$\mu$m point source removed sky brightness and reddening maps 
($\Psi_{\mathrm{I100}}$, $\Psi_{\mathrm{I60}}$,  $\Psi_{\mathrm{S100}}$ and
$\Psi_{\mathrm{SEBV}}$, respectively). The higher the value of $\Psi$ the 
higher the intensity excess of the loop over the background, 
i.e. $\Psi$ can be used as a quality indicator. It was a requirement, that a 
loop in the catalogue must show $\Psi$\,$\ge$\,3 in at least one of the $\Psi$
values. Examples of the intensity profiles of 
high and low significance loops are shown in Fig.~A.1.

\item {\bf Color index:}
$\Delta I_{\mathrm{60}}/\Delta I_{\mathrm{100}}$ color indices 
were calculated for our loops from the radially avergaged 60 and 
100\,$\mu$m surface brightness profile. The color index is defined 
as the slope of the $I_{\mathrm{60}}$ versus $I_{\mathrm{100}}$ 
scatter plot using the data points of the surface brightness profiles 
in the positions of the loop wall only ($a_{\mathrm{in}}\,\le\,a\,\le\,a_{\mathrm{out}}$). 
The uncertainties of $\Delta I_{\mathrm{60}}/\Delta I_{\mathrm{100}}$ given in the catalogue 
are the formal errors of the slope fitting.

\item {\bf Associated objects:}
We attributed associated objects to our identified loops with the condition that
an associated object has to be placed in the wall or in the interior of the loop. 
We considered the following type of possible associated objects (references are indicated):

\begin{itemize}   
\item[$\bullet$] dark clouds (Dutra \& Bica \cite{Dutra02});   
\item[$\bullet$] supernova remnants (Green \cite{Green94});   
\item[$\bullet$] OB-associations (Lang \cite{Lang92});   
\item[$\bullet$] pulsars (Taylor et al. 1993);  
\item[$\bullet$] HII regions (Sharpless \cite{Sharpless59});   
\item[$\bullet$] IRAS point source with molecular core FIR colours;   
\item[$\bullet$] IRAS point source with T Tau star-like FIR colours.  
\end{itemize}

Selection criteria for molecular cores and T Tauri stars were selected  from the
IRAS Point Source Catalogue (Joint IRAS Science Working Group 1988)  according to
the following criteria:

\begin{itemize}   
\item[$\bullet$] point sources associated with galaxies were excluded; 
\item[$\bullet$] photometric qualities are 2 or better at 12, 25 and 60\,$\mu$m;  
\item[$\bullet$] 
 \begin{itemize}   \small 
  \item[$\circ$] molecular cores:

	    $0.4\le\log_{\mathrm{10}}\big({{F_{\mathrm{25}}}\over{F_{\mathrm{12}}}}\big)\le1.0$  \&
	    $0.4\le\log_{\mathrm{10}}\big({{F_{\mathrm{60}}}\over{F_{\mathrm{25}}}}\big)\le1.3$

   \item[$\circ$] T Tauri stars:

	    $0.0\le\log_{\mathrm{10}}\big({{F_{\mathrm{25}}}\over{F_{\mathrm{12}}}}\big)\le0.5$  \&
            $-0.2\le\log_{\mathrm{10}}\big({{F_{\mathrm{60}}}\over{F_{\mathrm{25}}}}\big)\le0.4$ 
 \normalsize
 \end{itemize}   
\end{itemize}      

\noindent following the definitions by Emerson (1998).  $F_{\mathrm{12}}$, $F_{\mathrm{25}}$ and
$F_{\mathrm{60}}$ are the 12, 25 and 60\,$\mu$m uncorrected IRAS fluxes, respectively.

\end{itemize}

\end{document}